\documentclass[preprint2]{aastex62}
\usepackage{color}
\usepackage[titletoc]{appendix}
\usepackage[fleqn]{amsmath}
\usepackage{float}
\usepackage{threeparttable}
\usepackage{comment}
\usepackage{enumitem}
\usepackage{natbib}

\newcommand{\unit}[1]{\ensuremath{\, \mathrm{#1}}}
\newcommand\kep{\textit{Kepler}}
\newcommand\gaia{\textit{Gaia}}

\setcitestyle{notesep={ }}

\shortauthors{Ca\~nas et al.}
\shorttitle{Kepler-730: A Hot Jupiter system with a close-in planet companion}

\begin{document} 

\title{Kepler-730: A hot Jupiter system with a close-in, transiting, Earth-sized planet} 
\correspondingauthor{Caleb I. Ca\~nas}
\email{canas@psu.edu}

\author{Caleb I. Ca\~nas}
\altaffiliation{NASA Earth and Space Science Fellow}
\affiliation{Department of Astronomy \& Astrophysics, The Pennsylvania State University, 525 Davey Lab, University Park, PA 16802, USA}
\affiliation{Center for Exoplanets \& Habitable Worlds, University Park, PA 16802, USA}
\affiliation{Penn State Astrobiology Research Center, University Park, PA 16802, USA}

\author{Songhu Wang} 
\affiliation{Department of Astronomy, Yale University, New Haven, CT 06511, USA}
\affiliation{\textit{51 Pegasi b} Fellow}

\author{Suvrath Mahadevan}
\affiliation{Department of Astronomy \& Astrophysics, The Pennsylvania State University, 525 Davey Lab, University Park, PA 16802, USA}
\affiliation{Center for Exoplanets \& Habitable Worlds, University Park, PA 16802, USA}
\affiliation{Penn State Astrobiology Research Center, University Park, PA 16802, USA}

\author{Chad F. Bender}
\affiliation{Department of Astronomy and Steward Observatory, University of Arizona, Tucson, AZ 85721, USA}

\author{Nathan De Lee}
\affiliation{Department of Physics, Geology, and Engineering Technology, Northern Kentucky University, Highland Heights, KY 41099, USA}
\affiliation{Department of Physics \& Astronomy, Vanderbilt University, Nashville, TN 37235, USA}

\author{Scott W. Fleming}
\affiliation{Space Telescope Science Institute, 3700 San Martin Drive, Baltimore, MD 21218, USA}

\author{D. A. Garc\'ia-Hern\'andez}
\affiliation{Instituto de Astrof\'isica de Canarias (IAC), E-38205 La Laguna, Tenerife, Spain}
\affiliation{Universidad de La Laguna (ULL), Departamento de Astrof\'isica, E-38206 La Laguna, Tenerife, Spain}

\author{Fred R. Hearty}
\affiliation{Department of Astronomy \& Astrophysics, The Pennsylvania State University, 525 Davey Lab, University Park, PA 16802, USA}

\author{Steven R. Majewski}
\affiliation{Department of Astronomy, University of Virginia, Charlottesville, VA 22904, USA}

\author{Alexandre Roman-Lopes}
\affiliation{Departamento de F\'isica, Facultad de Ciencias, Universidad de La Serena, Cisternas 1200, La Serena, Chile}

\author{Donald P. Schneider}
\affiliation{Department of Astronomy \& Astrophysics, The Pennsylvania State University, 525 Davey Lab, University Park, PA 16802, USA}
\affiliation{Center for Exoplanets \& Habitable Worlds, University Park, PA 16802, USA}

\author{Keivan G. Stassun}
\affiliation{Department of Physics \& Astronomy, Vanderbilt University, Nashville, TN 37235}
\begin{abstract}
Kepler-730 is a planetary system hosting a statistically validated hot Jupiter in a 6.49 day orbit and an additional transiting candidate in a 2.85 day orbit. We use spectroscopic radial velocities from the APOGEE-2N instrument, Robo-AO contrast curves, and \gaia{} distance estimates to statistically validate the planetary nature of the additional Earth-sized candidate. We perform astrophysical false positive probability calculations for the candidate using the available \kep{} data and bolster the statistical validation using radial velocity data to exclude a family of possible binary star solutions. Using a radius estimate for the primary star derived from stellar models, we compute radii of \(1.100^{+0.047}_{-0.050}\unit{R_{Jup}}\) and \(0.140\pm0.012\unit{R_{Jup}}\) (\(1.57\pm0.13\unit{R_{\oplus}}\)) for Kepler-730b and Kepler-730c, respectively. Kepler-730 is only the second compact system hosting a hot Jupiter with an inner, transiting planet.
\end{abstract}
\section{Introduction}
The formation pathways of hot Jupiter planets remains an active area of research \citep[see][and references therein]{Dawson2018}. Current theoretical paradigms for producing these behemoths fall largely into the following two main categories.
\newline
1). Dynamical migration (e.g., planet-planet scattering, \citealt{Rasio1996}; Lidov-Kozai cycling with tidal friction, \citealt{Wu2003}; and secular interactions, \citealt{Wu2011, Petrovich2015}) violently delivers giant planets to their current orbits, and leaves them dynamically hotter and isolated.
\newline
2). Hot Jupiters might alternatively be formed via quiescent disk migration \citep{Lin1996} or \textit{in situ} formation \citep{Batygin2016}, processes that leave the system dynamically cooler and compact.

Although the presence or absence of additional low-mass planets in close orbital proximity to hot Jupiters provides a zeroth-order test of distinct and competing formation mechanisms, the true occurrence rate for close-in planetary companions to systems with a hot Jupiter remains unclear \citep{Millholland2017, Wang2018}.

The radial velocity (RV) precision required for detecting companions with masses comparable to super-Earths, believed to be the most common type of planets in our Galaxy, are generally at or below $1-2\,{\rm m\,s^{-1}}$, a detection threshold achieved with the most precise spectrographs \citep{Fischer2016}. Transits by these planets cause drops in stellar brightness smaller than $\sim 0.1\%$, which remain beyond the capabilities of the current generation of wide-field ground-based transit surveys (see \citealt{Pepper2018} and references therein).

Hidden planets have started to emerge as higher photometric precision observations of existing planetary systems are obtained. WASP-47b is a typical hot Jupiter that was originally detected with SuperWASP \citep{Hellier2012}. Two additional transiting short-period super-Earths in the system were not detected until subsequent observations were obtained from the \textit{Kepler} spacecraft \citep{Borucki2010} during the \textit{K2} mission \citep{Becker2015}. Until recently, WASP-47 was the only confirmed hot Jupiter system known with additional close-in planet companions.

\citet{Thompson2018} used all four years of the \textit{Kepler} data to reveal another potential WASP-47-like system, Kepler-730, with a previously known hot Jupiter and an additional transiting planet candidate (also noted by \citealt{Zhu2018}). This object appears to be an Earth-sized inner planet with an orbital period of $2.85\,{\rm days}$, and was not detected in previous searches \citep{Steffen2012, Huang2016}.

In this Letter, we statistically validate the planetary nature of Kepler-730c based on Doppler velocimetry from the Sloan Digital Sky Survey (SDSS)/APOGEE-2 spectra, Robo-AO high-contrast imaging, and  \kep{} photometry.
\section{Observations and Data Reduction}
\subsection{APOGEE-2 Radial Velocities}\label{radsec}
Kepler-730 (KOI-929, KIC 9141746, 2MASS J19021315+4534438, $Kp$ = 15.65, $H$ = 14.18) was observed from the Apache Point Observatory (APO) between 2017 May 6 and 2018 June 21 as part of the APO Galaxy Evolution Experiment (APOGEE) program \citep{Majewski2017,Zasowski2017} to spectroscopically monitor a substantial fraction of the Kepler objects of Interest (KOIs; \citealt{Fleming2015}) as part of the SDSS-IV survey \citep{Blanton2017}. We obtained 16 spectra using the high-resolution (\(R\sim22,500\)), near-infrared (\(1.514-1.696\) \unit{\mu m}), multi-object APOGEE-2N spectrograph (\citealt{Wilson2012}, 2018 submitted to PASP), mounted on the Sloan 2.5m telescope \citep{Gunn2006}. 
\startlongtable
\begin{deluxetable*}{cccc}
\tablecaption{APOGEE-2 Observations\label{tab:table1}}
\tablehead{\colhead{BJD\(_\text{TDB}^{*}\)} & \colhead{RV\(^{\dagger}\) (km s\(^{-1}\))} & \colhead{\(1\sigma\) (km s\(^{-1}\))} & \colhead{S/N\(^\ddagger\)} (pixel\(^{-1}\))}
\startdata
2457879.872229 & -68.96 & 0.28 & 15 \\
2457908.783554 & -69.51 & 0.48 & 10 \\
2457918.770521 & -69.28 & 0.28 & 13 \\
2457919.788244 & -69.89 & 0.43 & 10 \\
2457920.797865 & -69.06 & 0.23 & 18 \\
2457938.765117 & -69.09 & 0.46 & 8 \\
2457940.716527 & -68.95 & 0.48 & 9 \\
2457941.715177 & -69.18 & 0.31 & 14 \\
2458007.742691 & -69.06 & 0.46 & 10 \\
2458188.013655 & -68.81 & 0.44 & 8 \\
2458209.986371 & -69.76 & 0.92 & 6 \\
2458234.981317 & -68.40 & 0.63 & 9 \\
2458237.922857 & -68.37 & 0.56 & 12 \\
2458238.917524 & -67.82 & 0.53 & 12 \\
2458261.832167 & -68.33 & 0.42 & 9 \\
2458290.776784 & -69.01 & 0.29 & 14 \\
\enddata
\tablenotetext{*}{BJD\(_\text{TDB}\) is the Barycentric Julian Date in the Barycentric Dynamical Time standard.}
\tablenotetext{\dagger}{The systemic velocity is \(\gamma=-68.94\unit{km\ s^{-1}}\).}
\tablenotetext{\ddagger}{APOGEE-2N has approximately two pixels per resolution element.}
\end{deluxetable*}
For each observation, the APOGEE-2 data  pipeline \citep{Nidever2015} performs sky subtraction, telluric and barycentric correction, and wavelength and flux calibration. We derived RVs using the maximum-likelihood cross-correlation method presented by \cite{Zucker2003}. We identified the best fitting synthetic spectrum in the $H$-band from a grid of BT-Settl synthetic spectra \citep{Allard2012} by cross-correlating the APOGEE spectrum with the highest signal-to-noise (S/N) against a grid spanning surface effective temperature (\(5300\le T_{e}\le5900\), in intervals of 100 K), surface gravity (\(3.5\le\log g\le4.5\), in intervals of 0.5 dex), metallicity (\(-0.5\le\text{[M/H]}\le0.5\), in intervals of 0.5 dex), and rotational broadening (\(2 \le v\sin i\le 50 \unit{km s^{-1}}\), in intervals of 2 \unit{km s^{-1}}). The synthetic spectrum with the largest correlation was then used for the final cross-correlation to derive the reported RVs and \(1\sigma\) uncertainties. These  values are listed in Table \ref{tab:table1}.
\subsection{Kepler Photometry}\label{sec:photreduc}
Kepler-730 was observed by \kep{} for a total of 15 quarters and has two planetary candidates, KOI-929.01 and KOI-929.02, with periods of \(\sim6.49\) days and \(\sim2.85\) days, respectively. KOI-929.01 (Kepler-730b)  was statistically validated as an exoplanet by \cite{Morton2016} with a false positive probability (FPP) for the signal of \(\lesssim 1\times10^{-4}\). Prior to the final \kep{} data release \citep[DR25;][]{Thompson2018}, KOI-929.02 was not considered a planetary candidate. 

For the purposes of statistical validation, we analyzed both the \kep{} simple aperture photometry (SAP) and pre-search data conditioned (PDCSAP) time-series light curves \citep{Stumpe2012} available at the Mikulski Archive for Space Telescopes (MAST). We detrended light curves using three methods: Cosine Filtering with Autocorrelation Minimization \citep[CoFiAM;][]{Kipping2013}, a polynomial analog of CoFiAM, and a Gaussian process. 

CoFiAM regresses the Kepler time series using a harmonic (or polynomial) series in a least-squares approach where the optimal detrending function is defined as the one that minimizes the autocorrelation of the residuals. For all of the detrending methods, the portion of the light curve within a factor of 0.6 of the transit duration (\(\pm0.6T_{14}\)) from each transit midpoint was excised prior to regression. For the polynomial and CoFiAM methods, each transit was processed separately using the data flanking half a period from each transit midpoint. A \(3\sigma\) clip on a 20-point rolling median was applied to the detrended light curve to remove any outliers. 

We used the {\tt celerite} package to perform the Gaussian process detrending, and assumed a quasi-periodic covariance function, following the procedure in \cite{Foreman-Mackey2017}.  Each quarter of data was detrended separately and no additional processing was done to the light curve. 

To prepare for statistical validation, the transits of the other planetary candidate were removed. The light curve was then phased to the period and time of conjunction listed in DR25 and trimmed to keep data within a phase of three of the transit duration (\(\pm3T_{14}\)). KOI-929.01 was detrended solely using a Gaussian process. KOI-929.02 was detrended using the three methods described above (CoFiAM, a polynomial basis, and a Gaussian process). The light curve for the joint fit presented in Section \ref{sec:fullfit} retained all of the data (including any overlapping transits) and each quarter was detrended using a Gaussian process.

\section{Planet Validation}
\subsection{Sky-projected Stellar Companions}
\begin{figure*}[!ht]
\epsscale{1.15}
\plotone{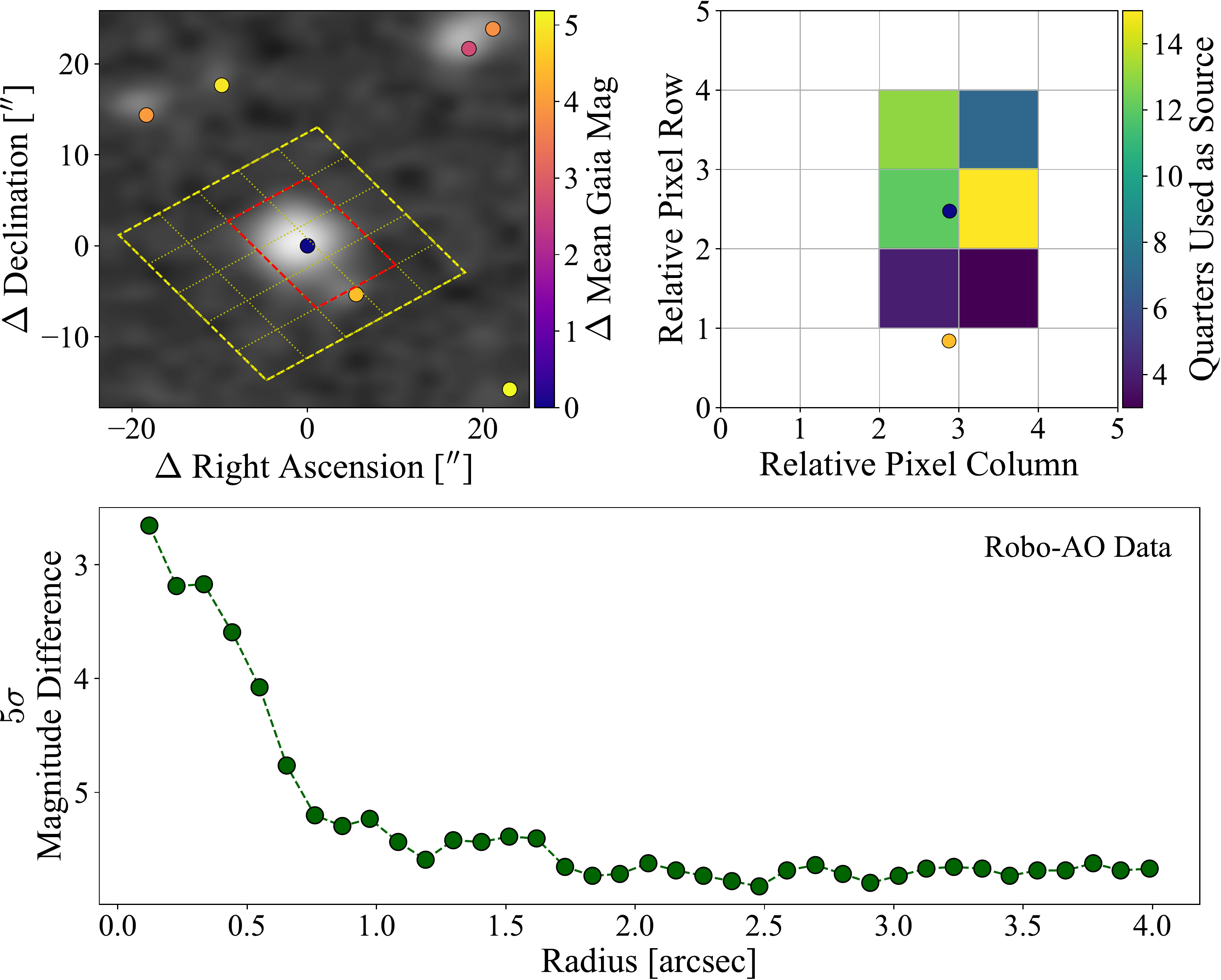}
\caption{Stellar background around Kepler-730. The upper left panel shows the six stars identified by \gaia{} within a sky-projected distance of \(30''\) atop an image of the same region from the Second Palomar Observatory Sky Survey (POSS-II/Red). The closest star, KIC 9141752, has a sky-projected distance of \(6.57''\). \gaia{} detected no other stars within the \(5\times5\) pixel mask (yellow grid) used by \kep{}. The \kep{} aperture mask (red grid) is highlighted in the upper right panel, with each pixel colored to the number of quarters it was used as a source pixel. Only Kepler-730 is contained in the aperture mask where the majority of the flux originates. The bottom panel shows the contrast curve provided by the Robo-AO survey illustrating the threshold magnitude difference to detect a stellar companion as a function of distance from Kepler-730. Robo-AO did not detect any other sources within \(4''\) of Kepler-730. \label{fig:f1}}
\end{figure*}

For the period range of these planetary candidates, the reliability of the \kep{} pipeline is \(>98\%\) \citep{Thompson2018}. There is also no other target in the \kep{} threshold crossing events that shares the same period as KOI-929.02, which suggests that the signal for this planetary candidate is unlikely to be produced by instrumental or stellar noise. The \kep{} photometry in MAST uses a \(5\times5\) pixel mask (see the upper row of Figure \ref{fig:f1}) to derive the light curves for this system, and each \kep{} pixel corresponds to \(\sim3.98''\). To investigate any potential background stars in the region, we used the latest data release from \gaia{} \citep{GaiaCollaboration2018} by searching a \(30''\) region around Kepler-730. A total of six stars reside within this region with the closest star, KIC 9141752 ($Kp$ = 19.1), located at a sky-projected distance of \(6.57''\). No other stars were located within the \kep{} pixel mask. In the upper right panel of Figure \ref{fig:f1}, the pixels that were considered source pixels varied by quarter, but rarely flanked the background star KIC 9141752. Even in the quarters with the smallest aperture mask, the transit of KOI-929.02 persisted.

The Robo-AO adaptive optics survey of the \kep{} field \citep{Ziegler2017} has observations of Kepler-730. The survey acquires images in an LP600 filter that serves to approximate the Kepler passband at redder wavelengths and mitigate the effects of blue wavelengths on instrumental performance. Robo-AO generates a contrast curve (bottom panel in Figure \ref{fig:f1}), providing the detection limit as a function of distance from Kepler-730; there are no detected companions within \(4''\). \cite{Ziegler2018} demonstrated that the recoverability of asterisms detected by Robo-AO in \gaia{} is \(\ge97\%\) for differences larger than three magnitudes at distances greater than \(2''\). While \gaia{} is often unable to resolve asterisms within \(\lesssim1''\) of a star, \gaia{} is more complete than Robo-AO for objects with mean \gaia{} magnitudes of \(G>20\). Together, the \gaia{} and Robo-AO data show that Kepler-730 has no close stellar companions within the \kep{} aperture mask. Figure \ref{fig:f1} displays the stars identified by \gaia{} within \(30''\), along with the \kep{} pixel mask and Robo-AO data.
\subsection{False Positive Analysis}
We adopted the {\tt vespa} package from \cite{Morton2016} to perform a false positive analysis of Kepler-730b. The algorithm validates a planet statistically by simulating and determining the likelihood of a range of astrophysical false positive scenarios that could generate the observed light curve. {\tt vespa} treats each planetary candidate as the only planet around the host star; this is a conservative view for Kepler-730 given the high reliability of \kep{} multiplanet systems \citep[e.g.,][]{Lissauer2014}. The code generates a population (20,000 systems) for each false positive scenario, including background eclipsing binaries (BEBs), eclipsing binaries (EBs), and hierarchical eclipsing binaries (HEBs), to calculate the likelihoods. We included the two artificial likelihood models from \cite{Morton2016} to flag if the transit signal did not fit any astrophysical model. The stellar properties for statistical validation were derived using the {\tt isochrones} package \citep{Morton2015} setting priors on the (i) 2MASS \(JHK\) magnitudes \citep{Skrutskie2006} and Kepler magnitudes, (ii) the \gaia{} parallax, (iii) the host star surface gravity, temperature and metallicity from the APOGEE Stellar Parameter and Chemical Abundances Pipeline \citep[ASPCAP;][]{GarciaPerez2016}, and (iv) the maximum visual extinction from estimates of Galactic dust extinction \citep[Bayestar17;][]{Green2018}. 

Two additional constraints for statistical analysis include the maximum radius permissible for a background eclipsing object and the maximum depth of the secondary transit. These values were adopted from \cite{Morton2016} for KOI-929.01. For KOI-929.02, the centroid offsets from the \kep{} data validation pipeline were used to determine the maximum radius. KOI-929.02 has centroid offsets of \(\sim1.5''\) and the maximum radius was set to a factor of three larger, at \(4.5''\). The maximum depth of the secondary was set to five times the uncertainty in the secondary depth from the \kep{} data validation pipeline. The Robo-AO contrast curve shown in Figure \ref{fig:f1} is an additional constraint applied to the {\tt vespa} analysis. 

The results of the statistical analyses for Kepler-730 are shown in Table \ref{tab:table2}. The light curve for KOI-929.01 was validated only using the PDCSAP flux, detrended with a Gaussian process, and has an FPP of \((1.7\pm1.4)\times10^{-4}\). The shallow transit depth of KOI-929.02 (\(\sim84\) ppm) warranted the use of different detrending mechanisms to determine its susceptibility to changes in detrending. For this candidate, we performed statistical validation on both the SAP and PDCSAP flux detrended with three methods described in Section \ref{sec:photreduc}. The values and respective errors for each analysis were calculated as the mean and standard deviation of a bootstrap of 10 iterations of {\tt vespa}. Regardless of the flux source and the detrending method, the signal was consistent with a statistically validated planet when adopting the threshold of FPP \(<1\%\) used in \cite{Morton2016}.
\startlongtable
\begin{deluxetable*}{cccccc}
\tablecaption{False Positive Probability Analysis of Kepler-730\label{tab:table2}}
\tablehead{\colhead{KOI} &
\colhead{FPP} &
\colhead{Source} &
\colhead{Polynomial} &
\colhead{CoFiAM} &
\colhead{Gaussian Process}
}
\startdata
929.02&All & SAP & \((9.1\pm2.4)\times10^{-5}\) & \((1.9\pm0.33)\times10^{-4}\) & \((8.1\pm2.5)\times10^{-5}\) \\
\(\cdots\)&Only EBs/HEBs & SAP & \((2.8\pm1.1)\times10^{-6}\) & \((2.6\pm0.64)\times10^{-7}\) & \((1.3\pm0.19)\times10^{-7}\) \\
\hline
929.01&All & PDCSAP & \(\cdots\) &\(\cdots\) & \((1.7\pm1.4)\times10^{-4}\) \\
\(\cdots\)&Only EBs/HEBs & PDCSAP & \(\cdots\) & \(\cdots\) & \((1.7\pm1.4)\times10^{-4}\) \\
929.02&All & PDCSAP & \((1.2\pm0.39)\times10^{-4}\) & \((1.2\pm0.34)\times10^{-4}\) & \((5.6\pm3.2)\times10^{-5}\) \\
\(\cdots\)&Only EBs/HEBs & PDCSAP & \((6.4\pm0.99)\times10^{-8}\) & \((5.7\pm1.3)\times10^{-8}\) & \((1.4\pm0.44)\times10^{-9}\) \\
\enddata
\end{deluxetable*}
\begin{figure*}[!ht]
\epsscale{1.15}
\plotone{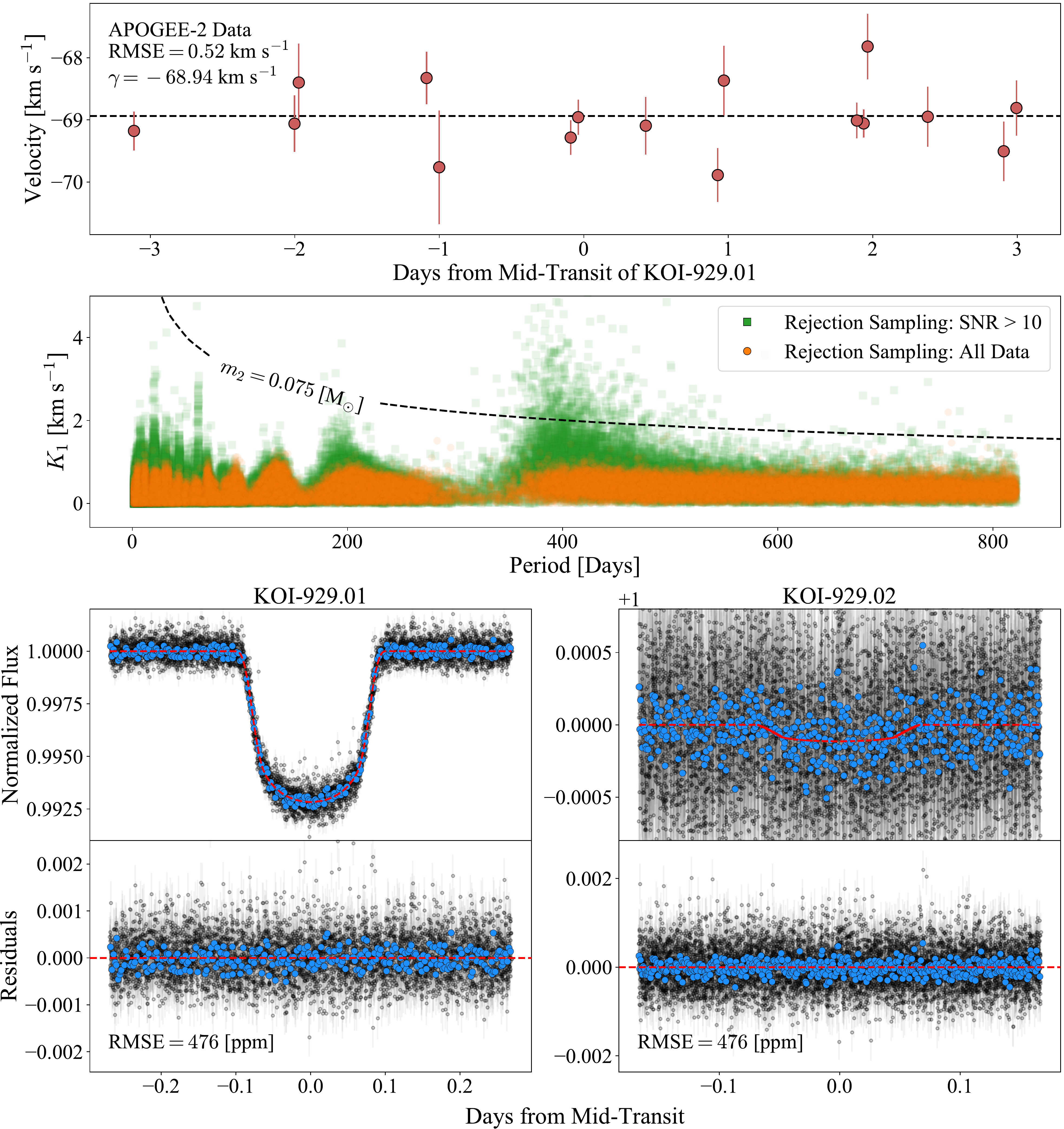}
\caption{Velocimetry and photometry of Kepler-730. The top panel shows the radial velocities phased to the period of KOI-929.01. The middle panel presents the surviving population with \(e<0.5\) after rejection sampling. The contour for a star at the hydrogen mass burning limit (with \(i=90^{\circ}\) and \(e=0\)) around Kepler-730 is plotted for reference. The code {\tt thejoker} performed \(>4\times10^6\) (\(2^{22}\)) samplings and the large surviving population (\(>60,000\)) demonstrates that our radial velocities are consistent with no statistically significant detection of a close stellar companion. The bottom row displays the phased light curves and best models with the rms error (RMSE). The small dots are the raw data and the larger circles are the data binned to a 1-minute cadence. \label{fig:f2}}
\end{figure*}
\subsection{RV Non-detection}
The derived RVs (Section \ref{radsec}) folded to the period of the hot Jupiter, KOI-929.01, are shown in the first panel of Figure \ref{fig:f2}. No physical solution exists when adopting a standard Keplerian orbit and maximizing the likelihood. To determine if our RVs supported the existence of any companion, we used {\tt thejoker} \citep{Price-Whelan2017} to perform a rejection sampling analysis on the APOGEE-2 data. We performed the same analysis on the entire RV data set and the subset derived from spectra with \(\text{S/N}> 10\) to determine if the quality of the data would mask a potential planet RV signal. We ran \(>4\times10^6\) (\(2^{22}\)) samples with {\tt thejoker} and more than 60,000 survived for each data set. The surviving samples are shown in the second panel of Figure \ref{fig:f2}. The underlying samples do not favor any orbital solutions between 1.5 days and twice the baseline of the APOGEE-2 observations (\(\sim411\) days). The smallest stellar companion, a star at the hydrogen mass burning limit (\(M_{2}=0.075\unit{M_{\odot}}\), \(i=90^{\circ}\), and \(e=0\)), would induce observable reflex motion of Kepler-730 with an amplitude of a few km s\(^{-1}\). {\tt vespa} does not use RVs in the statistical analysis. Instead, a non-detection in RVs can bolster the statistical validation by reducing or eliminating the contribution of HEB/EB scenarios. The non-detection was most significant for the hot Jupiter, KOI-929.01, where the probability that the transit signal is not due to EBs or HEBs was \(<10^{-6}\). These low false positive probabilities suggest that Kepler-730 is, statistically, a multiplanet system. 
\section{System Parameters}\label{sec:fullfit}
We used the {\tt EXOFASTv2} analysis package \citep{Eastman2017} to model the photometry. The priors included the (i) 2MASS \(JHK\) magnitudes, (ii) \(UBV\) magnitudes \citep{Everett2012}, (iii) Wide-field Infrared Survey Explorer magnitudes \citep{Wright2010}, (iv) spectroscopic parameters from ASPCAP, (v) maximum visual extinction from estimates of Galactic dust extinction from Bayestar17, and (vi) the distance estimate from \cite{Bailer-Jones2018}. The spectroscopic parameters are derived from a combined spectrum, are empirically calibrated, and have been determined to be reliable \citep[see][]{Holtzman2018}. The composite spectrum has a S/N \(\approx53\) per pixel and provides the following: \(T_{e}=5595\pm135\) K, \(\log g\sim4.06\), and \(\text{[Fe/H]}=0.21\pm0.02\). The surface gravity was poorly constrained during the calibration step and is only an initializing value for our analysis. Each planet had its period and time of mid-transit fixed to the value derived in DR25. The bottom row of Figure \ref{fig:f2} presents the result of the fit to the photometry and Table \ref{tab:table3} provides a summary of the stellar priors together with the inferred system parameters and respective confidence intervals.
\startlongtable
\begin{deluxetable*}{llcc}
\tablecaption{Parameters for the Kepler-730 System \label{tab:table3}}
\tablehead{\colhead{~~~Parameter} &
\colhead{Units} &
\multicolumn{2}{c}{Median Value}
}
\startdata
\sidehead{Primary Stellar Priors:}
~~~Effective Temperature\(^\dagger\) \dotfill & $T_{e}$ (K)\dotfill & \multicolumn{2}{c}{$5595 \pm 135$}\\
~~~Surface Gravity\(^\dagger\) \dotfill & $\log(g_1)$ (cgs)\dotfill & \multicolumn{2}{c}{$4.06$}\\
~~~Metallicity\(^\dagger\) \dotfill & [Fe/H]\dotfill & \multicolumn{2}{c}{$0.21\pm 0.02$}\\
~~~Maximum Visual Extinction \dotfill & \(A_{V,max}\) \dotfill & \multicolumn{2}{c}{$0.126$}\\
~~~Distance\dotfill & (pc)\dotfill & \multicolumn{2}{c}{$1935 \pm 122$}\\
\sidehead{Primary Parameters:}
~~~Mass \dotfill & $M_{1}$ (\unit{M_{\odot}})\dotfill & \multicolumn{2}{c}{$1.047^{+0.072}_{-0.054}$}\\
~~~Radius\dotfill & $R_{1}$ (\unit{R_{\odot}})\dotfill & \multicolumn{2}{c}{$1.411^{+0.049}_{-0.051}$}\\
~~~Density \dotfill & $\rho_1$ (g \unit{cm^{-3}})\dotfill & \multicolumn{2}{c}{$0.529^{+0.057}_{-0.046}$}\\
~~~Surface Gravity \dotfill & $\log(g_1)$ (cgs)\dotfill & \multicolumn{2}{c}{$4.162^{+0.032}_{-0.028}$}\\
~~~Effective Temperature\dotfill & $T_{e}$ (K)\dotfill & \multicolumn{2}{c}{$5620^{+55}_{-59}$}\\
~~~Metallicity\dotfill & [Fe/H]\dotfill & \multicolumn{2}{c}{$0.210\pm0.014$}\\
~~~Age\dotfill & (Gyr)\dotfill & \multicolumn{2}{c}{$9.5^{+2.5}_{-2.7}$}\\
~~~Parallax\dotfill & (mas)\dotfill & \multicolumn{2}{c}{$0.495^{+0.020}_{-0.019}$}\\
~~~Linear Limb-darkening Coefficient\dotfill & $u_1$\dotfill & \multicolumn{2}{c}{$0.418^{+0.028}_{-0.029}$}\\
~~~Quadratic Limb-darkening Coefficient\dotfill & $u_2$\dotfill & \multicolumn{2}{c}{$0.235\pm0.045$}\\
\multicolumn{2}{l}{Planetary Parameters\(^\ddagger\):} & b & c\\
~~~Orbital Period\dotfill & $P$ (days) \dotfill&6.491682808&2.851883380\\
~~~Time of Mid-transit\dotfill & $T_C$ (BJD\textsubscript{TDB})\dotfill & 2455007.633553 & 2454965.145500\\
~~~Scaled Radius\dotfill & $R_{p}/R_{1}$ \dotfill & $0.08013^{+0.00074}_{-0.00084}$&$0.01025\pm0.00074$\\
~~~Radius\dotfill & $R_{p}$  (\unit{R_{Jup}}) \dotfill& $1.100^{+0.047}_{-0.050}$&$0.140\pm0.012$\\
~~~Scaled Semi-major Axis\dotfill & $a/R_{1}$ \dotfill & $10.60^{+0.38}_{-0.32}$&$6.10^{+0.21}_{-0.18}$\\
~~~Semi-major Axis\dotfill & $a$ (AU) \dotfill& $0.0694^{+0.0016}_{-0.0012}$&$0.03997^{+0.00089}_{-0.00069}$\\
~~~Orbital Inclination\dotfill & $i$ (degrees)\dotfill & $86.96^{+0.37}_{-0.31}$&$83.81^{+1.10}_{-0.83}$\\
~~~Impact Parameter\dotfill & $b$\dotfill & $0.561^{+0.038}_{-0.050}$&$0.659^{+0.079}_{-0.110}$\\
~~~Transit Duration\dotfill & $T_{14}$ (hours)\dotfill & $4.33\pm0.03$ & $2.76\pm0.26$\\
~~~Equilibrium Temperature\dotfill & $T_{eq}$ (K)\dotfill & $1219^{+21}_{-22}$&$1607^{+27}_{-29}$\\
\enddata
\tablenotetext{\dagger}{Values from ASPCAP.}
\tablenotetext{\ddagger}{\(P\) and \(T_{C}\) are fixed to the \kep{} values. \(e\) and \(\omega\) are null.}
\end{deluxetable*}
The modeling reveals that Kepler-730 is a subgiant star with a radius of $1.411^{+0.049}_{-0.051}$ \unit{R_{\odot}}. It hosts a hot Jupiter and an interior Earth-sized planet with radii of $1.100^{+0.047}_{-0.050}$ \unit{R_{Jup}} and $0.140\pm0.012$ \unit{R_{Jup}} (\(1.57\pm0.13\unit{R_{\oplus}}\)), respectively. To ensure that the derived parameters were consistent, we applied the diagnostic explored in \cite{Seager2003} for a transiting system and proceeded to estimate the primary stellar density from the photometry to be \(0.537_{-0.048}^{+0.063}\) and \(0.531_{-0.046}^{+0.060}\) \unit{g\ cm^{-3}} for KOI-929.01 and KOI-929.02, respectively. These values are consistent with each other and are in agreement with the density derived from the stellar models listed in Table \ref{tab:table3}. For comparison, the density of KIC 9141752 derived from stellar models is \(3.48\pm0.45\) \unit{g\ cm^{-3}}.

The impact parameters also set a lower limit for the mutual inclination at \(\sim3^\circ\). To investigate if a system hosting 1\unit{M_{\oplus}} and 1\unit{M_{Jup}} planets could exist in this configuration, we performed an $N$-body simulation with {\tt whfast} \citep{Rein2015} spanning \(\sim 500\) Myr. While we ignore forces other than gravity and any effects from stellar evolution, the fact that both planets survived a long time suggests that a small mutual inclination does not necessitate chaotic evolution.
\section{Discussion}
The majority of currently detected hot Jupiters have no known close-in companions. The  WASP-47 system was, until recently, the only known exception. In this Letter, we validated a second such system, Kepler-730, which hosts a hot Jupiter with an inner, transiting planet, and sheds new light on the origins of hot Jupiters. The analysis of \gaia{}, Robo-AO, \kep{}, and APOGEE-2 data have revealed that the observed transits have a very high statistical probability of being genuine planets, and as such, provides independent validation of both Kepler-730b and Kepler-730c. The similar stellar densities derived from each transit further reinforces this conclusion. The \kep{} transit timing observations catalog \citep{Holczer2016} detected no timing variations, making it difficult to constrain the planetary masses. The non-detection of a Keplerian orbit in the APOGEE-2 velocimetry places an upper limit on the mass of the hot Jupiter of \(\lesssim13\unit{M_{Jup}}\), corresponding to a \(3\sigma\) detection.

The existence of close-in companions in hot Jupiter systems is possible evidence that precludes a dynamically violent history. The measurement of stellar obliquity for the Kepler-730 system thus provides an unique chance to test if spin-orbit misalignment of hot Jupiters is a natural consequence of high-eccentricity migration. From the derived system parameters, we predict that the semi-amplitude of the Rossiter-Mclaughlin effect for Kepler-730b is $\sim12\unit{m\ s^{-1}}$ (assuming \(v\sin i\sim2\unit{km\ s^{-1}}\)), which is marginally measurable with Keck/HIRES given the faintness of Kepler-730 \citep{Wang2018a}.

While tempting to discuss occurrence rates of such systems, we note that a significant fraction of the \kep{} hot and warm Jupiter sample has yet to be confirmed or statistically validated \citep{Huang2016}. Without additional observations, such as velocimetry, high-contrast imaging, and measured stellar parameters, a genuine false positive scenario can appear to be a statistically validated planet \citep[e.g.,][]{Canas2018}. Our ongoing APOGEE-2 survey of KOIs will help investigate a significant fraction of this hot Jupiter sample, enabling a more accurate estimation of occurrence rates of WASP-47-like systems.

\section{Acknowledgements}
We thank Jon Jenkins, Christopher Spalding, and Wei Zhu for helpful discussion.
S.W. thanks the Heising-Simons Foundation for their generous support.
C.I.C. acknowledges support by NASA Headquarters under the NASA Earth and Space Science Fellowship Program - Grant 80NSSC18K1114. C.I.C., C.F.B., and S.M. acknowledge support from NSF award AST 1517592.
S.R.M. acknowledges support from NSF award AST 1616636.
D.A.G.H. acknowledges support provided by the Spanish Ministry of Economy and Competitiveness (MINECO) under grant AYA-2017-88254-P.

Some of the data presented in this Letter were obtained from MAST. STScI is operated by the Association of Universities for Research in Astronomy, Inc., under NASA contract NAS5-26555. Support for MAST for non-HST data is provided by the NASA Office of Space Science via grant NNX09AF08G and by other grants and contracts. 2MASS is a joint project of the University of Massachusetts and IPAC at Caltech, funded by NASA and the NSF.
DSS was produced at STScI under USG grant NAG W-2166. Images of these surveys are based on photographic data obtained using the OST on Palomar Mountain and the UKST. The plates were processed into the present compressed digital form with the permission of these institutions.
Funding for the \kep{} mission is provided by the NASA Science Mission directorate. The NASA Exoplanet Archive is operated by Caltech, under contract with NASA under the Exoplanet Exploration Program.

Funding for SDSS-IV has been provided by the Alfred P. Sloan Foundation, the U.S. Department of Energy Office of Science, and the Participating Institutions. SDSS-IV acknowledges support and resources from the Center for High-Performance Computing at the University of Utah. The SDSS website is \url{www.sdss.org}. SDSS-IV is managed by the Astrophysical Research Consortium for the Participating Institutions of the SDSS Collaboration including the Brazilian Participation Group, the Carnegie Institution for Science, Carnegie Mellon University, the Chilean Participation Group, the French Participation Group, Harvard-Smithsonian Center for Astrophysics, Instituto de Astrof\'isica de Canarias, The Johns Hopkins University, Kavli Institute for the Physics and Mathematics of the Universe (IPMU) / University of Tokyo, Lawrence Berkeley National Laboratory, Leibniz Institut f\"ur Astrophysik Potsdam (AIP), Max-Planck-Institut f\"ur Astronomie (MPIA Heidelberg), Max-Planck-Institut f\"ur Astrophysik (MPA Garching), Max-Planck-Institut f\"ur Extraterrestrische Physik (MPE), National Astronomical Observatories of China, New Mexico State University, New York University, University of Notre Dame, Observat\'ario Nacional / MCTI, The Ohio State University, Pennsylvania State University, Shanghai Astronomical Observatory, United Kingdom Participation Group, Universidad Nacional Aut\'onoma de M\'exico, University of Arizona, University of Colorado Boulder, University of Oxford, University of Portsmouth, University of Utah, University of Virginia, University of Washington, University of Wisconsin, Vanderbilt University, and Yale University.

\end{document}